\documentclass[10pt,conference, letterpaper]{IEEEtran}
\usepackage[ansinew]{inputenc}
\usepackage{epsfig}
\usepackage[T1]{fontenc}
\usepackage{amsmath,enumerate,amssymb,hhline,verbatim}
\usepackage{xcolor}
\usepackage[square, comma, numbers]{natbib}

\newcommand{\bbar}[1]{\setbox0=\hbox{$#1$}\dimen0=.2\ht0 \kern\dimen0 \overline{\kern-\dimen0 #1}}

\newcommand{\Q}{\mathbb{Q}}
\newcommand{\C}{\mathbb{C}}
\newcommand{\R}{\mathbb{R}}

\newcommand{\ZZ}{\mathbb{Z}}

\providecommand{\abs}[1]{\ensuremath{\left\lvert #1 \right\rvert}}

\providecommand{\norm}[1]{\ensuremath{\left\Vert #1 \right\Vert}}

\providecommand{\vv}[1]{\textquotedblleft #1\textquotedblright}

\DeclareMathOperator*{\Vol}{Vol}
\renewcommand{\IEEEQED}{\IEEEQEDopen}

\newtheorem{thm}{Theorem}[section]
\newtheorem{theorem}[thm]{Theorem}

\newtheorem{corollary}[thm]{Corollary}
\newtheorem{lem}[thm]{Lemma}
\newtheorem{lemma}[thm]{Lemma}
\newtheorem{proposition}[thm]{Proposition}
\newtheorem{prop}[thm]{Proposition}

\newtheorem{definition}[thm]{Definition}

\newtheorem{remark}[thm]{Remark}

\newcommand{\D}{{\mathcal D}}
\newcommand{\HH}{{\mathbf H}}
\newcommand{\mindet}[1]{\hbox{\rm det}_{min}\left( #1\right)}
\newcommand{\diag}{{\rm diag}}

\begin{document}
\title{Division algebra codes achieve MIMO block fading channel capacity within a constant gap}
\author{ 
\IEEEauthorblockN{Laura Luzzi}
\IEEEauthorblockA{Laboratoire ETIS,
 CNRS - ENSEA - UCP \\
Cergy-Pontoise, France \\
laura.luzzi@ensea.fr}
\and
\IEEEauthorblockN{Roope Vehkalahti}
\IEEEauthorblockA{Department of Mathematics and Statistics,
University of Turku\\
Finland\\
roiive@utu.fi}
}
\maketitle

\begin{abstract}
This work addresses the question of achieving capacity with lattice codes in multi-antenna block fading channels when the number of fading  blocks tends to infinity.
In contrast to the standard approach in the literature which employs random lattice ensembles,
 the existence results in this paper are derived  from number theory. 
It is shown that a multiblock construction based on division algebras achieves rates within a constant gap from block fading capacity both under maximum likelihood decoding and naive lattice decoding.  
First the gap to capacity is shown to depend on the discriminant of the chosen division algebra; then class field theory is applied to build families of algebras with small discriminants. The key element in the construction is the choice of a sequence of division algebras whose centers are number fields with small root discriminants. 
\end{abstract}

\section{Introduction}
The closed-form expression of the capacity of ergodic multiple-input multiple-output (MIMO) channels was given in \cite{FG} and \cite{Tel}. It was shown that this capacity is achievable within  a constant gap by combining simple modulation and a strong outer code \cite{HoBr}.   In this work we consider the question of achieving MIMO capacity with lattice codes and we prove that there exists  a family of 
so-called multi-block division algebra codes \cite{YB07, Lu} that achieve a constant gap to capacity over the block fading MIMO channel when the number of fading blocks tends to infinity.

Our  constructions are based on two results from classical class field theory. First  we choose the center $K$ of the algebra from an ensemble of Hilbert class fields having small root discriminant  and then we prove the existence of a $K$-central division algebra with small discriminant. Our lattices   belong to a very general family of division algebra codes introduced in \cite{YB07, Lu, EK}, and developed further in  \cite{HL} and \cite{VHO}. We will use the most general form presented in \cite{LSV}.

While we discuss specific lattice codes from division algebras, our proofs do work for any ensemble of  matrix lattices with asymptotically good normalized minimum determinant.  The larger this value is, the smaller the gap to the capacity.

This work suggests  that capacity questions in fading channels are naturally linked to problems in the mathematical research area of \emph{geometry of numbers}. 
Unlike our previous work in the single antenna case \cite{VL2014}, many of the questions that arise have not been actively studied by the mathematical community. 

We note that, while studying diversity-multiplexing gain tradeoff (DMT) of  multiblock codes in \cite{Lu},  H.-f. Lu   conjectured that these   codes might approach MIMO capacity. Our work confirms that conjecture; however, we point out that it is unlikely that DMT-optimality alone is enough to approach capacity. Instead one should pick the code very carefully by maximizing the normalized minimum determinant.

\section{Preliminaries}

\subsection{Channel model}
We consider a MIMO system with $n$ transmit and $n_r$ receive antennas, where transmission takes place over $k$ quasi-static Rayleigh fading blocks of  delay $T=n$. 
Each multi-block codeword $X\in M_{n\times nk}(\C)$ has the form $(X_1,X_2,\dots, X_k)$, where the submatrix $X_i \in M_n(\C)$ is sent during the $i$-th block. The received signals are given by
\begin{equation}\label{eq:channel}
Y_i=H_i X_i +W_i, \quad \quad i \in \{1,\ldots,k\}
\end{equation}
where $H_i \in M_{n_r \times n}(\C)$ and $W_i\in M_{n_r \times T}(\C)$ are the channel and noise matrices. The coefficients of $H_i$ and $W_i$ are modeled as circular symmetric complex Gaussian with zero mean and unit variance per complex dimension, and the fading blocks $H_i$ are independent. We assume that perfect channel state information is available at the receiver, and that decoding is performed after all $k$ blocks have been received.  We will call such a channel an $(n, n_r, k)$-multiblock channel. 

A multi-block code  $\mathcal{C}$ in a $(n, n_r, k)$-channel is  a set of matrices  in $M_{n\times nk}(\C)$.  In particular we will concentrate on finite codes that are drawn from lattices. Let $R$ denote the code rate in bits per complex channel use; equivalently, $\abs{\mathcal{C}}=2^{Rkn^2}$. 
We assume that every matrix $X$ in a finite code $\mathcal{C}\subset M_{n\times nk}(\C)$ satisfies the average power constraint
\begin{equation} \label{power_constraint_multiblock}
\frac{1}{nk} \norm{X}^2 \leq P.
\end{equation}

\subsection{Lattices and spherical shaping }

\begin{definition}
A {\em matrix lattice} $L \subseteq M_{n\times T}(\C)$ has the form
$
L=\ZZ B_1\oplus \ZZ B_2\oplus \cdots \oplus \ZZ B_s,
$
where the matrices $B_1,\dots, B_s$ are linearly independent over $\R$, i.e., form a lattice basis, and $s$ is
called the \emph{rank}  or the \emph{dimension} of the lattice.
\end{definition}

In the following we will use the notations $\R(L)$ for the linear space which is generated by the basis elements of the lattice $L$, and $\Vol(L)$ for the volume of a fundamental region of $L$
according to the Lebesgue measure in $\R(L)$.
\begin{lem}\emph{\cite{GL}}\label{shift}
Let us suppose that $L$ is a lattice in  $M_{n\times kn}(\C)$  and  $S$ is a Jordan measurable bounded subset of $\R(L)$.  Then there exists $X \in M_{n\times kn}(\C)$ such that
$$
|(L+X)\cap S|\geq\frac{\mathrm{Vol}(S)}{\mathrm{Vol}(L)}.
$$
\end{lem}

\subsection{Minimum determinant for the multiblock channel}\label{multiblockdesign}
Let us first assume that we have an $l$-dimensional square matrix lattice $L$ in  $M_{n\times n}(\C)$. The minimum determinant of the lattice $L$ is defined as
$$
\mindet{L}=\inf_{X \neq \{\bf 0\}} \{|\det(X)|\}.
$$
The pairwise-error probability based determinant criterion by Tarokh \emph{et al.} \cite{TSC} motivates us to define the {\em normalized minimum determinant} $\delta(L)$, which  is obtained  by scaling the lattice $L$ to have a unit volume fundamental parallelotope 
before taking the minimum determinant.
 A simple computation proves the following:
\begin{lemma}\label{scale}
Let $L \subset M_{n\times n}(\C)$ be an $l$-dimensional matrix lattice. We then have that
$$
\delta(L) =\mindet{L}/(\Vol(L))^{n/l}.
$$
\end{lemma}

This concept generalizes to the multiblock case as follows.

Let us suppose that $L\subset M_{n\times kn}(\C)$  is a multiblock code and that $X=(X_1,X_2,\dots X_k)$ is a codeword in $L$.
The received signal matrix
$$
(H_1X_1, H_2X_2,\dots, H_k X_k) +(W_1,W_2,\dots, W_k),
$$
can  just as well be written in the form
$$
(H_1,H_2,\cdots, H_k)\diag(X) +\diag(W_1, W_2,\dots, W_k),
$$
where the diag-operator places the $i$-th $n\times n$ entry in the $i$-th diagonal block of a matrix
in $M_{kn\times kn}(\C)$. This reveals that  optimizing a code $L$ for the $(n, n_r, k)$-multiblock channel is equivalent to optimizing
$\diag(L)$ for the usual one shot $nk\times kn_r$ MIMO channel, where $\diag(L)$ is defined as $\{\diag(X)\mid X\in L\}$.
\begin{definition}
By abusing notation we define
$$
\mindet{L}:=\mindet{\diag(L)}\,\,\,\mathrm{and}\, \,\,\delta(L):=\delta(\diag(L)).
$$
\end{definition}

\section{Lattices from division algebras }\label{oneshotalgebra}
Let us now describe how lattice codes from division algebras are typically built.

\begin{definition}\label{cyclic}
Let $K$ be an algebraic number field of degree $m$ and assume   that $E/K$ is a cyclic Galois
 extension of degree $n$ with Galois group
$Gal(E/K)=\left\langle \sigma\right\rangle$. We can define an associative $K$-algebra
$$
\mathcal{D}=(E/K,\sigma,\gamma)=E\oplus uE\oplus u^2E\oplus\cdots\oplus u^{n-1}E,
$$
where   $u\in\mathcal{D}$ is an auxiliary
generating element subject to the relations
$xu=u\sigma(x)$ for all $x\in E$ and $u^n=\gamma\in K^*$. We call the resulting algebra a \emph{cyclic algebra}.
\end{definition}

It is clear that the center of the algebra $\mathcal{D}$ is precisely the field $K$. That is, an element of $\D$ commutes with all other elements of $\D$ if and only if it lies in $K$.

\begin{definition}
 We  call $\sqrt{[\D:K]}$ the \emph{degree} of the algebra $\D$. It is easily verified that the degree of $\D$ is equal to $n$.
\end{definition}

\begin{definition}
 A \emph{$\ZZ$-order} $\Lambda$ in $\D$ is a subring of $\D$ having the same identity element as
$\D$, and such that $\Lambda$ is a finitely generated
module over $\ZZ$ which generates $\mathcal{D}$ as a linear space over $\Q$.
\end{definition}

To every $\ZZ$-order $\Lambda$ in $\D$ we can associate a non-zero integer $d(\Lambda/\ZZ)$ called the \emph{$\ZZ$-discriminant of $\Lambda$}
\cite[Chapter 2]{Reiner}. 

\section{Asymptotically good families of division algebra codes}\label{construction}

\subsection{Number fields with small root discriminants} \label{Martinet_section} 
The following theorem by Martinet \cite{Martinet} proves the existence of infinite sequences of number fields $K$ with small discriminants $d_K$. As we will see, choosing such a field as the center of the algebra $\mathcal{D}$ is the key to obtaining a good normalized minimum determinant.   

\begin{theorem} \label{Martinet_theorem}
There exists an infinite tower of totally complex number fields $\{K_k\}$ of degree $2k=5\cdot2^t$, such that
\begin{equation} \label{G}
 \abs{d_{K_k}}^{\frac{1}{2k}}=G,
\end{equation}
for $G \approx 92.368$, and an infinite tower of totally real number fields $\{F_k\}$ of degree $k=2^t$ such that 
\begin{equation} \label{G1}
 \abs{d_{F_k}}^{\frac{1}{k}}=G_1,
\end{equation}
where $G_1 \approx 1058$. 
\end{theorem}

\subsection{Division algebra based $2kn^2$-dimensional codes  in $M_{n\times nk}(\C)$}\label{highdegree}
Let us now suppose that we have a totally complex  field  $K$ of degree $2k$ and a $K$-central division algebra $\D$ of degree  $n$. 
\begin{proposition}\cite{LSV}\label{complex}
Let  $\Lambda$ be a $\ZZ$-order in $\D$. Then there exists an injective mapping $\psi:\D\mapsto M_{n\times nk}(\C)$ such that $\psi(\Lambda)$ is a $2kn^2$-dimensional lattice in $M_{n\times nk}(\C)$ and 
\begin{align*}
& \mindet{\psi(\Lambda)}= 1, \,\, \,\, \Vol(\psi(\Lambda))= \sqrt{|d(\Lambda/\ZZ)|\cdot 2^{-2kn^2}}, \\
& \delta(\psi(\Lambda))=\left(\frac{2^{2kn^2}}{|d(\Lambda/\ZZ)|}\right)^{1/4n}.
\end{align*}
\end{proposition}
We can now see that in order to maximize the minimum determinant of a  multiblock code, we should minimize the $\ZZ$-discriminant of the corresponding $\mathcal{O}_K$-order $\Lambda$, 
given by
$$d(\Lambda/\mathbb Z)=N_{K/\mathbb Q}(d(\Lambda/\mathcal O_K))(d_K)^{n^2},$$
where 
$N_{K/\mathbb Q}$ is the algebraic norm in $K$.

Let 
$P_1$ and $P_2$ be some prime ideals of $K$ with norms $N_{K/\Q}(P_1)$ and $N_{K/\Q}(P_2)$. 
According to \cite[Theorem 6.14]{VHLR} there exists  a degree $n$ division algebra $\D$ having $\ZZ$-order $\Lambda$ with discriminant
\begin{equation}\label{discriminant2}
d(\Lambda/\ZZ)=(N_{K/\mathbb Q}(P_1) N_{K/\mathbb Q}(P_2))^{n(n-1)} (d_K)^{n^2}.
\end{equation}

Let us now aim for building the families of $(n,n, k)$ multiblock codes with as large as  possible normalized minimum determinant.

A trivial observation is that every number field  of degree $2k$ has prime ideals $P_1$ and $P_2$ such that
\begin{equation}\label{trivial}
N_{K/\mathbb Q}(P_1)\leq 2^{2k} \,\,\mathrm{and}\,\, N_{K/\mathbb Q}(P_2) \leq 3^{2k}.
\end{equation}
Armed with this observation, we have the following.
\begin{proposition} \label{prop:volume}
Given $n$ there exists a family of $2n^2k$-dimensional lattices $L_{n,k}\subset M_{n\times nk}(\C)$, 
such  that
\begin{align*}
&{\det}_{min}(L_{n,k})=1, \quad \Vol(L_{n,k}) \leq 6^{kn(n-1)}\left(\frac{G}{2}\right)^{n^2k}.
\end{align*}
\end{proposition}
\begin{IEEEproof}
Suppose that $K$ is a degree $2k$ field extension in the Martinet family of totally complex fields such that (\ref{G}) holds. We know that this field $K$ has some primes $P_1$ and $P_2$  such that $N_{K/\mathbb Q}(P_1)\leq 2^{2k} \,\,\mathrm{and}\,\,N_{K/\mathbb Q}(P_2) \leq 3^{2k}$. Then, there exists a central division algebra $\mathcal{D}$ of degree $n$ over $K$, and a maximal order $\Lambda$ of $\mathcal{D}$, such that
\begin{align}
&d(\Lambda/\mathbb{Z})=(N_{k/\mathbb Q}(P_1) N_{K/\mathbb Q}(P_2))^{n(n-1)} (d_K)^{n^2} \leq \notag \\
& \leq (6^{2k})^{n(n-1)} (G^{2k})^{n^2}. \tag*{\IEEEQED}%
\end{align}
\let\IEEEQED\relax%
\end{IEEEproof}%
\vspace{-3ex}
Let us now show how we can design multiblock codes $\mathcal{C}$ having rate
 $R$, and satisfying average power constraint $P$, from a scaled version $\alpha L_{n,k}\subseteq M_{n\times nk}(\C)$ of the lattices defined in Proposition \ref{prop:volume}. Here $\alpha$ is a suitable energy normalization constant. We denote by $B(r)$ the set of matrices in $M_{n \times nk}(\C)$ with Frobenius norm smaller or equal to $r$. According to Lemma \ref{shift}, we can choose a constant shift $X_R\in M_{n\times nk}(\C)$ such that for $\mathcal{C}=B(\sqrt{Pkn} )\cap(X_R+\alpha L_{n,k})$ we have
$$2^{Rnk}= \abs{\mathcal{C}}\geq\frac{\Vol(B(\sqrt{Pkn} ))}{\Vol(\alpha L_{n,k})}=\frac{C_{n,k}P^{n^2k}}{\alpha^{2n^2k} \Vol(L_{n,k})},$$
where $C_{n,k}=\frac{(\pi n k)^{n^2 k}}{(n^2 k)!}$. We find the following condition for the scaling constant:
\begin{equation} \label{alpha_multiblock}
\alpha^2=\frac{C_{n,k}^{\frac{1}{n^2k}} P}{2^{\frac{R}{n}} \Vol(L_{n,k})^{\frac{1}{n^2 k}}} \geq \frac{C_{n,k}^{\frac{1}{n^2k}} P}{2^{\frac{R}{n}} (G/2) 6^{1-\frac{1}{n}}}.
\end{equation}

\subsection{Division algebra based $kn^2$-dimensional codes  in $M_{n\times nk}(\C)$}\label{lowdegree}

Let $K/\Q$ be a totally real  number field of degree $k$ and $\D$ a $K$-central division algebra of degree $n$. Then there 
 exists an embedding $\psi:\D\rightarrow M_{n\times nk}(\C)$ \cite{VHO} with the following properties. 

\begin{proposition}\label{quaternion}
Let us suppose that $\Lambda$ is a $\ZZ$-order in $\D$.
Then $\psi(\Lambda)$ is an $n^2k$-dimensional lattice in $M_{n\times nk}(\C)$ and
\begin{align*}
&\mindet{\psi(\Lambda)}= 1, \,\,\, \Vol(\psi(\Lambda))=\sqrt{d(\Lambda/\ZZ)}, \,\, \mathrm{ and } \\
&\delta({\psi(\Lambda)})=\left(\frac{1}{|d(\Lambda/\ZZ)|}\right)^{1/2n}.
\end{align*}
\end{proposition}
Next, we will focus on a particular  instance of this 
family of lattices. We 
use the notation $\HH$ for matrices of the form 
$$
\begin{pmatrix}
c& -b^*\\
b&c^*
\end{pmatrix}
$$
and $M_{1\times k}(\HH)$ for matrices in $M_{2\times 2k}(\C)$ such that each $2\times 2$ block is of type $\HH$.

\begin{proposition}\cite{Alamouti_MISO}\label{alamouti}
Let $K$ be a totally real number field of degree $k$. Then there exists a degree $2$ $K$-central division algebra $\D$ and a $\ZZ$-order $\Lambda \subset \D$  such 
that $\psi(\Lambda)$ is a $4k$-dimensional lattice in $M_{1\times k}(\HH)\subseteq M_{2\times 2k}(\C)$ and
$$
\mindet{\psi(\Lambda)}= 1, \,\,\, \Vol(\psi(\Lambda))=|d_K|^2. 
$$
\end{proposition}

Assuming that the center $K$ belongs to the family of real fields from Theorem \ref{Martinet_theorem}, we have the following.
\begin{corollary}\label{alamouti2}
For every   $k=2^m$, 
there exists a $4k$-dimensional lattice $L_{Alam,k}\subset M_{1\times k}(\HH)$  such that
$$
\mindet{\psi(\Lambda)}= 1, \,\,\, \Vol(\psi(\Lambda))=G_1^{2k},
$$
where $G_1\approx 1058$.
\end{corollary}

Similarly to the previous section, we will produce codes $\mathcal{C}$ having rate $R$ and satisfying average power constraint $P$ from the lattices
$L_{Alam,k}\subseteq M_{2\times 2k}(\C)$. Let $\alpha$ be an energy normalization constant. According to  Lemma \ref{shift}, 
$\exists X_R\in M_{2\times 2k}(\C)$ such that for $\mathcal{C}=B(\sqrt{2Pk} )\cap( X_R+\alpha L_{Alam,k})$ we have
$$2^{2Rk}=|\mathcal{C}| \geq\frac{\Vol(B(\sqrt{Pk2} ))}{\Vol(\alpha L_{Alam,k})}=\frac{C_{Alam,k}P^{2k}}{\alpha^{4k} \Vol(L_{Alam,k})},$$
where $C_{Alam,k}=\frac{{(2k\pi)}^{2k}}{(2k)!}$. Solving for 
$\alpha$, we find 
\begin{equation} \label{alpha_multiblock_alam}
\alpha^2=\frac{C_{Alam,k}^{\frac{1}{2k}} P}{2^{R} \Vol(L_{n,k})^{\frac{1}{2k}}} \geq \frac{C_{Alam,k}^{\frac{1}{2k}} P}{2^{R}G_1  }. 
\end{equation}

\section{Achieving constant gap}

\subsection{The codes for the $(n,n,k)$-multiblock channel}
Let us now consider the lattice codes $\mathcal{C}$ of section \ref{highdegree}. Here the underlying lattice $L_{n,k}\subset M_{n \times nk}(\C)$ is $2n^2k$-dimensional. We are considering the channel model \eqref{eq:channel} in the symmetric MIMO case where $n_r=n$. We will analyze the performance of these codes  when the number of antennas $n$ is fixed and the number of blocks $k$ tends to infinity. 
\begin{prop} \label{prop_highdegree}
Over the $(n,n,k)$ multiblock channel, reliable communication is guaranteed when $k \to \infty$ for rates
\begin{equation*}
R < n\left(\log\frac{P}{n}  e^{\frac{1}{n}\sum_{i=1}^n \psi(i)} -\log n  + \log\frac{\pi e}{2}  - \log 6^{1-\frac{1}{n}} G \right)
\end{equation*}
when using the multiblock code construction in Section \ref{highdegree}.
\end{prop}

\begin{remark} We can compare the achievable rate with the tight lower bound in \cite[eq. (7)]{ONBP_2002} for $n$ transmit and receive antennas\footnote{We note that the capacity (per channel use) of the block fading MIMO channel of finite block length $T$ with perfect channel state information at the receiver is independent of $T$ \cite[eq. (9)]{Marzetta_Hochwald}. So the bounds in \cite{Tel} and \cite{ONBP_2002} still hold in our case.}:
$$C \geq n \log\left(1+\frac{P}{n}e^{\frac{1}{n}\sum_{i=1}^n \psi(i)}\right)$$
\end{remark}

\begin{IEEEproof}[Proof of Proposition \ref{prop_highdegree}]
Let $d_H$ denote the minimum Euclidean distance in the received constellation:
$$d_{H}^2=\min_{\substack{X, \bar{X} \in \mathcal{C}\\ X \neq \bar{X}}} \sum_{i=1}^k \norm{H_i (X_i-\bar{X}_i)}^2.$$
Suppose that the receiver performs maximum likelihood decoding or \vv{naive} lattice decoding (closest point search in the infinite 
lattice). 
For both, the error probability is bounded by 
$$P_e \leq \mathbb{P}\left\{ \norm{W}^2 \geq \left(\frac{d_{H}}{2}\right)^2\right\},$$ 
where $W=(W_1,\ldots,W_k)$ is the multiblock noise. Note that
{\allowdisplaybreaks
\begin{align*}
&d_{H}^2 \geq \alpha^2 n  \min _{X \in L_{n,k} \setminus \{0\}} \sum_{i=1}^k \abs{\det(H_iX_i)}^{\frac{2}{n}} \geq \\
& 
\hspace{-1ex}\geq \alpha^2nk \min _{X \in L_{n,k} \setminus \{0\}} \prod_{i=1}^k \abs{\det(H_iX_i)}^{\frac{2}{nk}} \geq \alpha^2 nk\prod_{i=1}^k \abs{\det(H_i)}^{\frac{2}{nk}}
\end{align*}
}%
where the  first step comes from  the Minkowski inequality,  the second step comes from the arithmetic mean - geometric mean 
inequality, and the third from observing that $\prod_{i=1}^k \abs{\det(X_i)}\geq 1$ for all $X \in L_{n,k} \setminus \{0\}$. Therefore 
\begin{equation*}
P_e \leq \mathbb{P}\left\{ \frac{\norm{W}^2}{kn^2} \geq \frac{\alpha^2}{4n} \prod_{i=1}^k \abs{\det(H_i)}^{\frac{2}{nk}}\right\}
\end{equation*}
Given $\epsilon>0$, we can bound the error probability by 
\begin{equation} \label{error_probability_multiblock}
\mathbb{P}\bigg\{ \frac{\norm{W}^2}{kn^2} \geq 1 + \epsilon\bigg \} + \mathbb{P}\bigg\{ \frac{\alpha^2}{4n}\prod_{i=1}^k \abs{\det(H_i)}^{\frac{2}{nk}} <1 + \epsilon\bigg\}  
\end{equation}
Note that $2\norm{W}^2 \sim \chi^2(2kn^2)$, and the tail of the chi-square distribution is bounded as follows for $\epsilon \in (0,1)$ \cite{Laurent_Massart}:
\begin{equation} \label{chi_square_bound}
\mathbb{P}\left\{\frac{\norm{W}^2}{kn^2} \geq 1 + \epsilon \right\} \leq 2 e^{-\frac{kn^2\epsilon^2}{8}}.
\end{equation}  
Therefore the first term in (\ref{error_probability_multiblock}) 
when $k \to \infty$. In order to upper bound the second term, we need to analyze the distribution of the random variable $\prod_{i=1}^k \abs{\det(H_i)}^{2}$. \\
In the single block case, it is well-known \cite{Goodman,Edelman} that if $H$ is an $n \times n$ matrix with i.i.d. complex Gaussian entries having variance per real dimension $1/2$, the random variable $2^n\abs{\det(H)}^2$
is distributed as the product $V_n=Z_1\cdots Z_n$ of $n$ independent chi square random variables $Z_j \sim \chi^2(2j), \; j \in \{1,\ldots,n\}$ with density $p_{Z_j}(x)=\frac{1}{2^j\Gamma(j)}x^{j-1}e^{-\frac{x}{2}}$. We have
$$\mathbb{E}[\ln Z_j]= \frac{1}{2^j\Gamma(j)} \int_0^{\infty}  x^{j-1} e^{-\frac{x}{2}} \ln x\; dx= \psi(j) +\ln 2,$$
where $\psi(x)$ is the Digamma function. Let 
\begin{equation} \label{M_n}
M_n=\mathbb{E}[\ln V_n] =n\ln 2 + \sum_{i=j}^n \psi(j).
\end{equation}
Observe that 
{\allowdisplaybreaks
\begin{align}
&\mathbb{E}[Z_j^{-v}]=\frac{1}{2^j\Gamma(i)}  \int_0^{\infty} x^{j-1-v} e^{-\frac{x}{2}} dx=\frac{\Gamma(j-v)}{2^v \Gamma(j)}, \label{eq1}\\
& \mathbb{E}[Z_j^{-v}\ln Z_j]=\frac{1}{2^j\Gamma(j)} \int_0^{\infty} x^{j-1-v} e^{-\frac{x}{2}} \ln x \; dx = \notag \\
&=\frac{\Gamma(j-v)}{2^v \Gamma(j)}(\psi(j-v)+\ln 2). \label{eq2}
\end{align} }
Now let's turn to the multiblock case, and let $S_k=2^{nk}\prod_{i=1}^k \abs{\det(H_i)}^2$. We have $S_k=V_n^{(1)}\cdots V_n^{(k)}$, where $V_n^{(i)}=Z_1^{(i)}\cdots Z_n^{(i)}$ are i.i.d. products of $n$ independent chi squared random variables $Z_j^{(i)} \sim \chi^2(2j)$. Note that $\mathbb{E}[\ln S_k]=k \mathbb{E}[\ln V_n]=k M_n$. Consider the zero-mean random variable
\begin{align*}
&B_k=-\ln S_k + kM_n=-\sum_{i=1}^k \ln V_n^{(i)} + kM_n=
\sum_{i=1}^k T_n^{(i)},
\end{align*}
where $T_n^{(i)}$ are i.i.d. with distribution 
$T_n=-\sum_{j=1}^n \ln Z_j + M_n$.
From the Chernoff bound \cite{Proakis} for $B_k$, given $\delta>0$, $\forall v>0$ we have
\begin{equation} \label{Chernoff_bound}
\mathbb{P}\left\{B_k \geq nk\delta\right\} \leq e^{-v\delta n k} \mathbb{E}[e^{vB_k}].
\end{equation}
The tightest bound in (\ref{Chernoff_bound}) is obtained for $v_{\delta}$ such that 
\begin{equation} \label{v_delta}
\mathbb{E}[B_k e^{v_{\delta} B_k}]= \delta n k \mathbb{E}[e^{v_{\delta} B_k}].
\end{equation}
It is easy to see that
\begin{align*}
&\mathbb{E}[e^{vT_n}]=e^{vM_n} \prod_{j=1}^n \mathbb{E}[Z_j^{-v}]=e^{v(M_n-n \ln 2)}\prod_{j=1}^n \frac{\Gamma(j-v)}{\Gamma(j)}
\end{align*}
Recalling that the variables $Z_j$ are independent, we find
{\allowdisplaybreaks
\begin{align*}
&\mathbb{E}[T_n e^{v T_n}]=\mathbb{E}\Big[\Big(-\sum_{i=j}^n \ln Z_j +M_n\Big)e^{vM_n} \Big(\prod_{l=1}^n Z_l^{-v}\Big)\Big]=\\
&=e^{vM_n} \Big( \sum_{j=1}^n\mathbb{E} [-Z_j^{-v} \ln Z_j] \prod_{l \neq j} \mathbb{E}[Z_l^{-v}] +M_n \prod_{l=1}^n\mathbb{E}[ Z_l^{-v}]\Big)=\\
&=e^{v(M_n-n\ln 2)} \Big(\prod_{l=1}^n \frac{\Gamma(l-v)}{\Gamma(l)}\Big)\Big(M_n-n \ln 2-\sum_{j=1}^n \psi(j-v)\Big).
\end{align*}
}
We can finally compute
\begin{align*}
&\mathbb{E}[e^{v B_k}]=
\mathbb{E}\left[ e^{v \sum_{i=1}^k T_n^{(i)}}\right]=\left(\mathbb{E}[e^{vT_n}]\right)^k=\\
&=e^{vk(M_n - n\ln2)}\prod_{j=1}^n \frac{\Gamma(j-v)^k}{\Gamma(j)^k}
\end{align*}
Similarly,
{\allowdisplaybreaks
\begin{align*}
&\mathbb{E}[B_k e^{v B_k}]=
\sum_{i=1}^k \mathbb{E} \left[ T_n^{(i)} e^{vT_n^{(i)}}\right] \mathbb{E}\left[ e^{\sum_{l \neq i} v T_n^{(l)}}\right]=\\
&=k e^{kv(M_n-n\ln2)} \Big(M_n -n\ln 2 -\sum_{j=1}^n \psi(j-v)\Big) \prod_{l=1}^n \frac{{\Gamma(l-v)}^k}{\Gamma(l)^k} 
\end{align*}
}
Thus, the tightest bound (\ref{v_delta}) is achieved for $v_\delta$ such that
\begin{equation*} 
n\delta=M_n - n\ln 2 -\sum_{j=1}^n \psi(j-v_{\delta})= \sum_{j=1}^n (\psi(j)-\psi(j-v_{\delta}))
\end{equation*}
Clearly, for fixed $n$, $v_{\delta} \to 0$ when $\delta \to 0$. From (\ref{Chernoff_bound}) we get
{\allowdisplaybreaks
\begin{align*}
&\mathbb{P}\{S_k^{\frac{1}{nk}} \leq e^{\frac{M_n}{n}-\delta}\}=\mathbb{P}\left\{B_k \geq nk \delta\right\} \leq \\
& \leq e^{k(v_{\delta}(-n\delta + M_n -n \ln 2)-\sum\limits_{j=1}^n (\ln\Gamma(j) - \ln\Gamma(j-v_{\delta})))}=\\
&=e^{k(\ln \Gamma(1-v_{\delta}) + v_{\delta} \psi(1-v_{\delta}) + \sum\limits_{i=2}^n (-\ln \Gamma(i) +\ln \Gamma(i-v_{\delta}) +v_{\delta}\psi(i-v_{\delta})))}.
\end{align*}
}%
Recall that $\Gamma(x)$ is monotone decreasing for $0 <x < a_0=1.461632\ldots$ and monotone increasing for $x>a_0$. Using the mean value theorem for the function $\ln \Gamma(x)$ in the interval $[i-v_{\delta},i]$ we get that for $i=1$, $v_{\delta} \psi(1-v_{\delta})+\ln \Gamma(1-v_{\delta}) \leq 0$, and for $i\geq 2$, $v_{\delta} \psi(i-v_{\delta}) \leq \ln \Gamma(i) - \ln \Gamma (i-v_{\delta})$. Thus, 
\begin{align*}
&\mathbb{P}\left\{2 \prod_{i=1}^k \abs{\det(H_i)}^{\frac{2}{nk}} \leq  e^{\frac{M_n}{n}-\delta}\right\} =\mathbb{P}\left\{S_k^{\frac{1}{nk}} \leq e^{\frac{M_n}{n}-\delta}\right\}\leq \\
&\leq e^{-k K_{n,\delta}}
\end{align*}
for some positive constant $K_{n,\delta}$. The second term in (\ref{error_probability_multiblock}) vanishes 
when $k \to \infty$ provided that $\frac{8n(1+\epsilon)}{\alpha^2} < e^{\frac{M_n}{n}-\delta}$. 
Recalling the bound for $\alpha$ from (\ref{alpha_multiblock}), a sufficient condition 
is 
$$\frac{4n(1+\epsilon)2^{\frac{R}{n}} 6^{1-\frac{1}{n}} G}{(C_{n,k})^{\frac{1}{n^2 k} P}} < e^{\frac{M_n}{n}-\delta}.$$

From Stirling's approximation, for large $k$ we have 
$(C_{n,k})^{\frac{1}{n^2k}} \approx \pi e/(n(2\pi n^2 k)^{\frac{1}{2n^2k}})$.  
Thus, we find that any rate
\begin{multline}
R < n\left(\log P -\frac{1}{2n^2 k} \log(2 \pi n^2 k) + \log\frac{\pi e}{4(1+\epsilon)} \right.+\\ \left. +\log (e^{\frac{M_n}{n}-\delta})  -2 \log n  - \log 6^{1-\frac{1}{n}} G\right), \label{rate_bound}
\end{multline}
where the logarithms are understood to be binary, is achievable asymptotically as $k \to \infty$. Note that
$e^{\frac{M_n}{n}}=e^{\ln 2 + \frac{1}{n} \sum_{i=1}^n \psi(i)}= 2 e^{\frac{1}{n}\sum_{i=1}^n \psi(i)}$. 
Since (\ref{rate_bound}) holds $\forall \delta>0, \forall \epsilon>0$, this concludes the proof.
\end{IEEEproof}

\begin{remark}
The number field towers we used are not the best known.  In fact there exists a family of totally complex fields such that $G<82.2$ \cite {Hajir_Maire}, but this would add some notational complications. Just as well the estimate given in  equation \eqref{trivial} is 
not optimal and it is likely that we can
reduce the term  $\log 6$ in the achievable rate formula.
\end{remark}

\subsection{The $(2,1,k)$-multiblock channel}
Let us now consider the codes of Corollary \ref{alamouti2} in the $2\times 1$ block fading channel. Here the matrices $H_i$ are simply vectors
$[h_1,h_2]$ and we suppose that the delay is 2.  The codewords in the lattice $L_{Alam, k}$ have block structure $X=[X_1,\dots,X_k]$, where 
each $X_i\in \HH$. \\
For these codes we can prove the following: 

\begin{prop} \label{prop_Alamouti}
Over the $(2,1,k)$ multiblock channel, reliable communication is guaranteed when $k \to \infty$ for 
$$R < \log\left(\frac{Pe^{1-\gamma}}{2}\right)+\log \frac{\pi e}{4} -\log G_1$$
when using the multiblock code construction $L_{Alam,k}$. 
\end{prop}
 
The proof is 
sketched in Appendix \ref{Appendix_Alamouti}.  We can compare the achievable rate in Proposition \ref{prop_Alamouti} to the tight lower bound on ergodic capacity in \cite[eq. (7)]{ONBP_2002} for $n=2$ and $n_r=1$:
$$C \geq \log\left(1+\frac{P}{2} e^{1-\gamma}\right).$$

\begin{small}

\end{small}
\newpage


\appendices


\section{Proof of Proposition \ref{prop_Alamouti}} \label{Appendix_Alamouti}
By direct calculation  $\norm{ H_iX_i}^2=\frac{1}{2}\norm{H_i}^2\norm{X_i}^2= \norm{H_i}^2 \abs{\det(X_i)}$. Using this result and the NVD property we have
$$d_{H}^2=\alpha^2 \min_{\substack{X \in L_{Alam,k} \setminus \{0\}}} \sum_{i=1}^k \norm{H_i X_i}^2 \geq \alpha^2 k \prod_{i=1}^k \norm{H_i}^{\frac{2}{k}}.$$
Therefore 
\begin{equation} \label{upper_bound2}
P_e \leq \mathbb{P}\left\{ \frac{\norm{W}^2}{2k} \geq \frac{\alpha^2}{8} \prod_{i=1}^k \norm{H_i}^{\frac{2}{k}}\right\}.
\end{equation}
Note that $2\norm{W}^2\sim \chi^2(4k)$ and $2\norm{H_i}^2 \sim \chi^2(4)$. From (\ref{upper_bound2}), we have that $\forall \epsilon>0$,
\begin{align*} 
P_e \leq \mathbb{P}\bigg\{ \frac{\norm{W}^2}{2k} \geq 1 + \epsilon\bigg \} + \mathbb{P}\bigg\{ \frac{\alpha^2}{8} \prod_{i=1}^k \norm{H_i}^{\frac{2}{k}} <1 + \epsilon\bigg\}
\end{align*}
When $k \to \infty$, the first term vanishes due to the bound (\ref{chi_square_bound}).

Let $S_k=2^k \prod_{i=1}^k \norm{H_i}^2=\prod_{i=1}^k Z_2^{(i)}$, where $Z_2^{(i)}$ are independent i.i.d. random variables with distribution $\chi^2(4)$. Let $\eta=\mathbb{E}[\ln Z_2]=1-\gamma+\ln 2$. The Chernoff bound for the zero-mean variable $B_k=-\ln S_k +k \eta$ implies that 
$$\mathbb{P}\{ B_k \geq k \delta\} \leq e^{-vk \delta} \mathbb{E}[ e^{vB_k}],$$
and for a fixed value of $\delta>0$, the tightest bound is given by $v_{\delta}$ such that $\mathbb{E}[B_k e^{vB_k}]=\delta k \mathbb{E}[e^{vB_k}]$. 
Recalling the identities (\ref{eq1}) and (\ref{eq2}), it is not hard to see that 
\begin{align*}
&\mathbb{E}[e^{vB_k}]=e^{vk(1-\gamma)} (\Gamma(2-v))^k,\\
&\mathbb{E}[B_k e^{vB_k}]=e^{vk(1-\gamma)} k (\Gamma(2-v))^k (1-\gamma - \psi(2-v))
\end{align*}
Thus, the optimal value $v_{\delta}$ is such that 
$$\delta=-\psi(2-v_{\delta})+1-\gamma,$$
and tends to $0$ when $\delta \to 0$. 
We can conclude that
\begin{align*}
& \mathbb{P}\left\{ B_k \geq k\delta\right\} \leq e^{-kv_{\delta}(\delta-1+\gamma)} (\Gamma(2-v))^k=\\
&= e^k (v_{\delta} \psi(2-v_{\delta}) + \ln \Gamma(2-v_{\delta})).
\end{align*}
The last exponent is negative when $v_{\delta}>0$ is small, since the mean value theorem for the function $x \mapsto \ln \Gamma(x)$ implies that 
$$-\ln\Gamma(2-v_{\delta}) =\ln \Gamma(2)-\ln \Gamma(2-v_{\delta}) \geq v_{\delta} \psi(2-v_{\delta}) .$$
Thus, $\forall \delta>0$,
\begin{align*}
& \mathbb{P}\{ B_k \geq k \delta\} = \mathbb{P}\{S_k \leq e^{-k(\delta-\eta)}\}=\\
&=\mathbb{P}\left\{ 2 \prod_{i=1}^k \norm{H_i}^{\frac{2}{k}} \leq e^{-(\delta-\eta)}\right\}=\\
&= \mathbb{P}\left\{\prod_{i=1}^k \norm{H_i}^{\frac{2}{k}} \leq e^{-(\delta-1+\gamma)}\right\} \to 0. 
\end{align*}
Thus, $\forall \delta>0$, the second term in the pairwise error probability bound vanishes provided that 
\begin{equation} \label{Alamouti_bound}
\alpha^2\geq \frac{8(1+\epsilon)}{e^{1-\gamma-\delta}},
\end{equation}
Comparing with (\ref{alpha_multiblock_alam}), we find the condition
$$\frac{8(1+\epsilon)}{e^{1-\gamma-\delta}} \leq \frac{C_{Alam,k}^{\frac{1}{2k}} P}{2^R G_1} \approx   \frac{\pi e P}{(4k \pi)^{\frac{1}{4k}} 2^R G_1},$$
where the last approximation holds for large $k$ due to Stirling's formula. Since the previous bound holds $\forall \epsilon>0, \forall \delta>0$, any rate 
$$R < \log\left(\frac{Pe^{1-\gamma}}{2}\right)+\log \frac{\pi e}{4} -\log G_1$$
is achievable. \hspace*{\fill}~\IEEEQED\par

\section{Geometry of numbers for fading channels}

In this paper we considered the problem of building capacity-approaching lattice codes for block fading multiple antenna channels.
One of the key elements of this approach was recognizing a geometric  invariant of the lattice which 
provides a design criterion. Let us now see how this approach fits into a more general context and can be regarded as a natural generalization of the classical theory of lattices for Gaussian channels.
Finally we show how the code design problems, both in Gaussian and fading channels, can be seen as instances of the same problem in the mathematical theory of geometry of numbers.

Consider a lattice $L \subset \C^n$ having fundamental parallelotope of volume one and define a function
$f_1:\C^n\to \R$ by 
\begin{equation}\label{euclidean}
f_1(x_1,\dots,x_n)=|x_1|^2+|x_2|^2+\cdots+ |x_n|^2.
\end{equation} 
The real number $h(L)=\inf_{x \in L, \, x\neq{\bf 0}} f_1(x)$ is  
called the
\emph{Hermite invariant} of the lattice $L$.  Let us now denote with
$\mathcal{L}_n$ the set of all $2n$-dimensional lattices in $\C^n$.

Suppose that we have an infinite family of lattices  $L_n\in \C^n$ with Hermite invariants satisfying $\frac{h(L_n)}{n}\geq c$, for some positive constant $c$. Then a classical result in information theory states that with this family of lattices, all rates satisfying
$$
R < \log_2(P) - \log_2\left(\frac{4}{c\pi e}\right),
$$
are achievable in the complex Gaussian channel \cite[Chapter 3]{CS}. This means that we can attach a single number $h(L_n)$ to each lattice $L_n\in \C^n$, which roughly describes its performance and in particular estimates how close to the capacity a family of lattices can get. This relation is one of the key connections between the theory of lattices and information theory \cite{CS} and has sparked a remarkable amount of research.

\smallskip

Let us now see how  our results in \cite{VL2014b} and in this paper can be seen as natural generalizations of the relation between Hermite constant and capacity.

Let us consider $2n$-dimensional lattices $L$ in $\C^n$ and the form
\begin{equation}\label{fastfading}
f_2(x_1,x_2,\dots, x_n)=|x_1x_2\cdots x_n|.
\end{equation}
Assuming again that we have a full lattice $L$, with $\Vol(L)=1$ 
we can define $\mathrm{Nd_{p, min}}(L)=\inf_{x\in L, x\neq{\bf 0}} f_2(x) $, the \emph{normalized product distance} of the lattice $L$.

Let us now assume that we have an infinite family of lattices  $L_n\in \C^n$ with normalized product distance satisfying $(\mathrm{Nd_{p, min}}(L_n))^{2/n}\geq c$, for some positive constant $c$. According to \cite{VL2014b} we then have that all rates satisfying
$$ R < \log_2(Pe^{-\gamma}) - \log_2\left(\frac{4}{\pi e}\right)+\log_2 c,$$
are accessible with this family of lattices with zero error probability in Rayleigh fast  fading channel.

\smallskip

Now consider a lattice  $L$ that lies in the space $M_{n\times kn}(\C)$. Let us suppose that $(X_1, X_2,\dots, X_k)$  is an element of $M_{n\times kn}(\C)$,  and define
$$
f_3(X_1, X_2,\dots, X_k)=\prod_{i=1}^k |\det(X_i)|.
$$
Assuming that $\Vol(L_k)=1$  we have  $$\delta(L)=\inf_{X\in L_k, X\neq{\bf 0}} f_3(X).$$

If we assume that we have a family of lattices $L_k\subset  M_{n\times kn}(\C)$ with the property that $\delta(L_k)^{2/kn}\geq c$ we have that any rate satisfying
$$
R < n\left(\log_2 \left(\frac{P}{n^2}  e^{\frac{1}{n}\sum_{i=1}^n \psi(i)}\right)  + \log_2\frac{\pi e}{2}   - \log_2 2+  \log c \right).
$$
is achievable with the lattices $L_k$.

We can now see that the normalized minimum determinant and product distance can be 
regarded
as generalizations of the Hermite invariant
which 
characterize
the gap to capacity achievable with a certain family of lattice codes. 
\smallskip

A natural question 
is how close to capacity we can get with these methods by taking the best possible lattice sequences.
Let us denote with $\mathcal{L}_{(n,k)}$ the set of all $2n^2k$-dimensional lattices in the space $M_{n \times nk}(\C)$ and formalize the question of achievable rates.

The \emph{Hermite constant} $H(k)$ can now be defined as 
\begin{equation}\label{f1}
H(2k)=\mathrm{sup}\{ h(L) \mid L \in \mathcal{L}_{(1,k)}, \Vol(L)=1\}.
\end{equation}
In the same manner we can define
\begin{equation}\label{f2}
\mathrm{Nd_{p, min}}(k)=\mathrm{sup}\{\mathrm{Nd_{p, min}}(L) \mid L \in \mathcal{L}_{(1,k)}, \Vol(L)=1\}.
\end{equation} 
and
\begin{equation}\label{f3}
\delta(k,n)=\mathrm{sup}\{\delta(L) \mid L \in \mathcal{L}_{(k,n)}, \Vol(L)=1\}.
\end{equation}
Each of these constants now represents how close to capacity our methods can take us. Any asymptotic lower bound with respect to $k$ will immediately provide a lower bound for the achievable rate. Just as well upper bounds will give upper bounds for the rates that are approachable with this method.

\smallskip

The questions of achievable rates have now been transformed into purely geometrical questions about existence of lattices with certain properties.
The value of the Hermite constant $H(k)$, for different values of $k$,  has been studied in mathematics for hundreds of years and there exists an extensive literature on the topic. In particular there exist good  upper and lower bounds and it has been proven that we can get quite close to Gaussian capacity with this approach \cite[Chapter 3]{CS}. \\
In the case of the product distance, the problem has been considered in the context of algebraic number fields and some upper bounds have been provided. 
As far as we know the best lower bounds come from the existence results provided by number field constructions \cite{Xing} and \cite{VL2014b}.

The 
properties 
of $\delta(k,n)$ 
are far less researched. Simple upperbounds can be derived from bounds for Hermite constants as pointed out in \cite{JV}
and lower bounds are 
obtained from division algebra constructions as described in this paper, but  
the mathematical literature doesn't seem to offer any ready-made results for this problem. \\
However, all three of these problems can be seen as special cases of a general problem in the mathematical theory of \emph{geometry of numbers} \cite{GLb}. Let us now elaborate on the topic.

\begin{definition}
A continuous 
function 
$F$: $M_{n\times kn}(\C) \to \R$
is called a homogeneous form of degree $\sigma>0$ if it satisfies the relation
$$
|F(\alpha {X})|=|\alpha|^{\sigma} |F(X)|\quad (\alpha \in \R, X \in M_{n\times kn}(\C)).
$$
\end{definition}

Let us consider the body $S(F)=\{X \,|\,X \in M_{n\times kn}(\C), |F(X)|\leq 1\}$, and a 
$2kn^2$ dimensional lattice $L$ with a  fundamental parallelotope of volume one.

We  then  define the \emph{homogeneous minima} $\lambda(F,L)$ of $F$ with respect to the lattice $L$ by
$$
\lambda(F,L )=(\mathrm{inf}\{\lambda|\,\lambda>0, \mathrm{dim}(\R(\lambda S(F)\cap L))\geq 1\})^{\sigma},
$$
where $\R(\lambda S(F)\cap L)$ is simply the $\R$-linear space generated by  the elements in $\lambda S(F)\cap L$.
This allows us to define the \emph{absolute homogeneous minimum} 
$$
\lambda(F)=\sup_{\Vol(L)=1}\lambda(F,L).
$$

We can now see that all of our forms $f_1$, $f_2$ and $f_3$ are homogeneous forms. For the  Hermite invariant we have $\sigma=2$, for the product distance $\sigma=n$, and for the normalized minimum determinant $\sigma=n^2k$. Easily we can also see that the constants \eqref{f1}, \eqref{f2} and \eqref{f3}
are just absolute homogeneous minima of the corresponding forms.

These results suggests that there is  a very general 
connection between
information theory and geometry of numbers in numerous channel models. 
It seems to be that given a fading channel model, there exists a form (or forms), whose absolute homogeneous minima 
provide 
a lower bound for the achievable rate 
using lattice codes.

\begin{remark}
The definitions for the geometry of numbers given in this section were stated for lattices in the space $M_{n\times nk}(\C)$, while normally the definitions are given in the space $\R^m$. This is however, just to keep 
our notation simple. 
The space $M_{n\times nk}(\C)$ can be very explicitly seen as the space $\R^{2n^2k}$ and we could have given the definitions also in the traditional form using this identification.

\end{remark}

\begin{small}
\bibliographystyle{chicago}  

\begin{thebibliography}{10}

\bibitem{FG} G. Foschini and M. Gans, ``On limits of wireless communications in a
fading environment when using multiple antennas'',\emph{ Wireless Personal
Communications}, March 1998.

\bibitem{Tel}
E.~Telatar, ``Capacity of multi-antenna {Gaussian} channels'', \emph{Europ.
  Trans. Telecomm.}, vol.~10, no.~6, pp. 585--595, 1999.

\bibitem{HoBr}  B.M. Hochwald, S. ten Brink, ``Achieving near-capacity on a multiple-antenna channel'', \emph{ {IEEE} Trans. Commun.} Vol.51,  Issue 3, 
pp.  389--399, March 2003.


\bibitem{YB07} S. Yang and J.-C. Belfiore,  ``Optimal space-time codes for the MIMO
amplify-and-forward cooperative channel'',  \emph{IEEE Trans. Inf. Theory},
vol. 53, no. 2, pp. 647--663, Feb. 2007.

\bibitem{Lu} H.-f. Lu, ``Constructions of multi-block space-time coding schemes that achieve the diversity-multiplexing tradeoff'',\emph{ IEEE Trans. Inform. Theory}, vol. 54, no. 8, pp. 3790-3796, Aug. 2008. 

\bibitem{EK} P. Elia, P. Vijay Kumar, ``Approximately-Universal Space-Time Codes for the Parallel, Multi-Block and
Cooperative-Dynamic-Decode-and-Forward Channels'',  available at http://arxiv.org/abs/0706.3502.

\bibitem{HL}
C.~Hollanti and H.-f. Lu, ``Construction methods for asymmetric and multi-block
  space-time codes,'' \emph{{IEEE} Trans. Inf. Theory}, vol.~55, no.~3, pp.
  1086 -- 1103, 2009.
  
  \bibitem{VHO} R. Vehkalahti, C. Hollanti, and F. Oggier, ``Fast-decodable asymmetric space-time codes from division algebras'', {\it IEEE Trans. Inf. Theory}, vol. 58, pp. 2362-- 2384, April 2012.
  



\bibitem{LSV} B. Linowitz, M. Satriano and R.Vehkalahti, ``A non-commutative analogue of the Odlyzko bounds and bounds on performance for space-time lattice codes'', available at http://arxiv.org/abs/1408.4630. 


\bibitem{VL2014} R. Vehkalahti and L. Luzzi, ``Number field lattices achieve Gaussian and Rayleigh channel capacity within a constant gap'', submitted to ISIT 2015, available at http://arxiv.org/abs/1411.4591.


\bibitem{GL}  P. M. Gruber and C. G. Lekkerkerker, \emph{Geometry of Numbers}, Elsevier, Amsterdam,
The Netherlands, 1987.


\bibitem{TSC} V. Tarokh, N. Seshadri, and A.R. Calderbank, ``Space-Time Codes for High Data Rate Wireless Communications: Performance Criterion and Code Construction'', {\it IEEE Trans. Inform. Theory}, vol. 44, pp. 744--765, March 1998.





\bibitem{Reiner} I. Reiner, {\it Maximal Orders}, Academic Press, New York 1975.


\bibitem{Martinet} J. Martinet, \vv{Tours de corps de classes et estimations de discriminants}, \emph{Invent. Math.} n. 44, 1978, pp. 65--73 

  \bibitem{VHLR}
R.~Vehkalahti, C.~Hollanti, J.~Lahtonen, and K.~Ranto, ``On the densest {MIMO}
  lattices from cyclic division algebras,'' \emph{{IEEE} Trans. Inform.
  Theory}, vol.~55, no.~8, pp. 3751--3780, Aug 2009.

\bibitem{Alamouti_MISO} R. Vehkalahti, \vv{Some properties of Alamouti-like MISO codes}, in \emph{IEEE Int. Symp. Inform. Theory}, Seoul, South Korea, 2009.

\bibitem{Laurent_Massart} B. Laurent, P. Massart, \vv{Adaptive estimation of a quadratic functional by model selection}, \emph{Annals of Statistics}, vol. 28, pp. 1302--1338, 2000.


\bibitem{Goodman} N. R. Goodman, \vv{The distribution of the determinant of a complex Wishart distributed matrix}, \emph{Ann. Math. Statist.}, vol. 34, pp. 178--180, 1963

\bibitem{Edelman} A. Edelman, \vv{Eigenvalues and condition numbers of random matrices}, Ph.D. Thesis, MIT 1989

\bibitem{Proakis} J. Proakis, \emph{Digital communications}, 4th ed., McGraw-Hill 2001


\bibitem{ONBP_2002} O. Oyman, R. Nabar, H. B\"olcskei, and A. Paulraj, \vv{Tight Lower Bounds on the Ergodic Capacity of Rayleigh Fading MIMO Channels}, 
\emph{IEEE GLOBECOM}, Nov.
2002, pp. 1172--1176.

\bibitem{Marzetta_Hochwald} T. L. Marzetta, B. M. Hochwald, \vv{Capacity of a mobile multiple-antenna communication link in Rayleigh flat fading}, \emph{{IEEE} Trans. Inform.
  Theory} vol 45 n.1, Jan 1999

\bibitem{Hajir_Maire} F. Hajir and C. Maire, \vv{Asymptotically good towers of global fields}, \emph{Proc. European Congress of Mathematics}, pp. 207--218, Birkh\"auser Basel, 2001.


%



%


%








\end{thebibliography}

\begin{thebibliography}{10}


\bibitem{CS} J.H. Conway and N.J.A. Sloane, {\it Sphere Packings, Lattices and Groups},
Springer-Verlag, New York, 1988.

\bibitem{VL2014b} R. Vehkalahti and L. Luzzi, ``Number field lattices achieve Gaussian and Rayleigh channel capacity within a constant gap'', submitted to ISIT 2015, available at http://arxiv.org/abs/1411.4591.

\bibitem{Xing} C. Xing,  ``Diagonal Lattice Space-Time Codes From Number Fields and Asymptotic Bounds'',  {\it IEEE Trans.  Inform. Theory}, vol.53, pp. 3921--3926,  November 2007.

\bibitem{JV} J. Lahtonen and R. Vehkalahti, ``Dense mimo matrix lattices - a meeting point for class field theory and invariant theory'', in Proc. Applied Algebra, Algebraic Algorithms, and Error Correcting Codes (AAECC-17), Bangalore, India, 2007.

\bibitem{GLb}  P. M. Gruber and C. G. Lekkerkerker, \emph{Geometry of Numbers}, Elsevier, Amsterdam,
The Netherlands, 1987.



\end{thebibliography}

\end{small}

\end{document}